\documentclass[preprint,review,12pt]{elsarticle}

\usepackage{cite}
\usepackage{amsmath,amssymb,amsfonts}
\usepackage{graphicx}
\usepackage{textcomp}
\usepackage{xcolor}
\usepackage{color,soul}
\usepackage{multirow}
\usepackage{url}
\usepackage{caption}
\usepackage{algorithm,algorithmic}
\usepackage{enumitem}
\usepackage{subcaption}
\usepackage{amsmath, amssymb}
\usepackage{setspace}

\journal{Electric Power Systems Research}

\begin{document}

\begin{frontmatter}


\title{Iterative Single-Loop Coordination of Transmission \\ and Distribution Systems and DER Aggregators \\ with Limited Information Sharing}

\author[KAUST,AGH]{Robert Sosnowski\corref{cor1}}
\ead{robert.sosnowski@kaust.edu.sa}
\author[JHU]{Yury Dvorkin}
\author[AGH]{Marcin Baszynski}
\author[KAUST]{Charalambos Konstantinou}
\cortext[cor1]{Corresponding author}

\affiliation[KAUST]{organization={King Abdullah University of Science and Technology},
            city={Thuwal},
            country={Kingdom of Saudi Arabia}}

\affiliation[AGH]{organization={AGH University of Krakow},
            city={Krakow},
            country={Poland}}

\affiliation[JHU]{organization={Johns Hopkins University},
            city={Baltimore},
            country={United States of America}}
            
\begin{abstract}
The growing integration of DERs underscores the importance of unified coordination across TSO, DSO, and aggregator levels.
This paper proposes a coordination method between TSO, DSOs and DER aggregators, based on a transparent, well-scalable, and decentralized single-loop iterative mechanism with limited information sharing. 
The approach enables iterative price-based coordination between adjacent levels, where the higher-level entities provide price signals and the lower-level ones respond with desired power schedules. 
The proposed method uses the proximal-gradient-based iterative co-optimization algorithm extended to a three-level case, which enjoys convergence characteristic of proximal algorithms. 
The performance of the proposed approach is compared with both a centralized formulation and a sequential reference formulation based on fixed transmission-level energy prices. The results demonstrate that the proposed coordination achieves outcomes very close to those obtained with the centralized formulation, while significantly improving performance relative to the sequential reference, particularly at higher levels of DER flexibility.
\end{abstract}



\begin{keyword}
DER \sep DSO \sep TSO \sep coordination \sep iterative \sep limited information sharing

\end{keyword}

\end{frontmatter}

\section*{Nomenclature}
\addcontentsline{toc}{section}{Nomenclature}

\begin{description}[leftmargin=!,labelwidth=24mm]
  \item[$t$] Time interval index
  \item[$i$, $j$] Bus index in the power flow formulations
  \item[$n$, $m$] Transmission and distribution system bus numbers
  \item[$g$, $T$, $D$, $A$] Index indicating generators, transmission, distribution or DER aggregators
  \item[$p$, $q$] Active and reactive power variables
  \item[$P$, $Q$, $S$] Active, reactive and apparent power constants
  \item[$v$, $\theta$, $l$] Bus voltage magnitude and angle, squared line current
  \item[$G$, $B$] Line series conductance and susceptance
  \item[$G^{sh}$, $B^{sh}$] Line shunt conductance and susceptance
  \item[$R$, $X$] Line resistance and reactance
  \item[$\psi$, $\delta$] Violations of the line power and bus voltage
  \item[$\kappa_{S/V}$] Power and voltage limits penalty coefficients
  \item[$\sigma$] Standard deviation of the power injection
  \item[$\gamma$] Deviation reactive-to-active power ratio
\end{description}

\section{Introduction}

The distributed energy resources (DERs) have various types of impacts on the electric power grid, among others technical, economic and environmental aspects~[\citenum{Marles}].
However, often only the influence on the distributed system is taken into account, while the impact on the transmission system is neglected~[\citenum{Gavgani}].
This limitation is reflected in the commonly applied sequential coordination scheme, in which transmission system prices are determined first, followed by coordination between the distribution system operators (DSOs) and DER aggregators (DERAs) based on these fixed prices.
To address this limitation, it is necessary to consider three-level coordination between the transmission system operator (TSO), DSOs and DERAs, as the coordination between these entities is inherently interdependent.

The need for simultaneous coordination across all system levels becomes particularly relevant when DERAs exhibit a high degree of flexibility and price responsiveness, as their aggregated reactions to price signals may significantly affect the power demand schedules of DSOs, and consequently, the power flows and marginal prices in the transmission system.
Typical examples of such highly flexible aggregators include energy storage systems and various types of responsive loads, such as EV charging infrastructure, smart buildings, and thermostatically controlled devices, including smart thermostats and electric water heaters.
The number of such resources is rapidly increasing: for instance, the global behind-the-meter battery energy storage capacity is expected to grow approximately sixfold between 2023 and 2030~[\citenum{IEA_bess}].
Similarly, the number of private home-installed light-duty vehicle charging points in the United States has increased nearly thirtyfold between 2015 and 2023~[\citenum{IEA_ev}].
Given the potentially large number of such aggregators in future power systems, the coordination mechanism must be scalable and decentralized. 
Moreover, minimizing information exchange is beneficial, as the involved entities are often independent and thus require coordination mechanisms that promote transparency and simplicity~[\citenum{Marques}].

The cost functions and constraints of all the participants in general may be relatively complex, especially if multiple interdependent time intervals, and both the active and reactive power markets are considered [\citenum{Capitanescu}]. 
For this reason, the standard bidding approach, in which participants submit their power capabilities and price offers in a single round, assuming a linear power-price relationship, is not sufficient to leverage the full potential of their cooperation.

On the other hand, including the cost functions and constraints of the lower-level entity in the formulation of the higher-level entity ensures an optimal solution.
However, this approach requires substantial disclosure of data, which is often undesirable due to confidentiality concerns and may reduce the transparency of the coordination process by limiting the autonomy of market participants [\citenum{Molzahn}].
Such extensive information disclosure is not only undesirable from the perspective of lower-level entities, as discussed above, but also from a higher-level perspective. Exchanging detailed cost functions and constraints significantly increases the communication burden, which becomes particularly critical in problems involving a large number of participants and multi-interval scheduling horizons, thereby reducing scalability. In practice, it is often the case that the communication infrastructure required for detailed asset-level coordination is not readily available or is fragmented across systems [\citenum{DOE_2020}]. Moreover, explicitly incorporating heterogeneous objective functions and constraints across system levels poses additional modeling and computational challenges and may require sophisticated optimization techniques. Therefore, coordination approaches based on limited and standardized information exchange are particularly attractive in such settings.
Leveraging the potential of coordination with limited information sharing is possible only through iterative methods. 
In such approaches, coordination emerges from the exchange of price signals and corresponding power schedules.

The considered problem can be formulated as a bi-level optimization problem, for which the coordination between DSOs and DERAs is performed in the inner iterative loop and the TSO and DSOs are coordinated in the outer one.
Such nested coordination structures have been proposed for hierarchical optimization problems involving different system levels, for example, in~[\citenum{WangQi}] for hierarchical optimal power flow calculations and in~[\citenum{Pandey2025}] for demand response frameworks.
However, employing a unified single-loop mechanism for TSO-DSOs-DERAs coordination can offer significant advantages, including reduced regulatory complexity, improved transparency, and a more coherent and scalable market structure.

Most of the approaches proposed in the literature that take into account the perspectives of TSOs, DSOs, and DERs (or DER aggregators) do not adopt iterative price-based coordination mechanisms with limited information sharing across both the TSO–DSOs and DSO–DERs interfaces.
For example, in [\citenum{Jiang, Chen, Mousavi}], the distribution market (DSO-DERs) is based on the bidding approach, in which the distributed generators provide their generation limits and the price bid once, without leveraging the variability of the price-power relations.
On the other hand, in approaches presented in [\citenum{Zhao, Gokcek, Najibi}], the information provided by DERs regarding the costs and constraints is directly incorporated in the DSO's optimization problem, which contradicts the principle of limited information sharing and may reduce the autonomy of market participants.

In other studies, in turn, important limitations arise from the design of their coordination mechanisms.
In~[\citenum{Li}], the TSO–DSOs coordination is based on nodal pricing using marginal prices, while prosumers trade energy and carbon emissions with each other and the DSO in a peer-to-peer (P2P) framework, modeled as a multi-leader multi-follower (MLMF) game. Such a combination of heterogeneous mechanisms increases regulatory complexity, reduces transparency, and leads to a more fragmented and less scalable market structure. 
The approach proposed in~[\citenum{Ullah}] employs a coherent coordination structure, but it relies on the alternating direction method of multipliers (ADMM), specifically a semi-proximal ADMM formulation, in which the higher-level entity shares not only price values with the lower-level entity, but also Lagrange multipliers and optimal power injections from its own perspective, which must be explicitly accounted for in the lower-level optimization problems.
Similarly, the framework proposed in~[\citenum{Zhai2025}], which considers microgrid networks as the lowest coordination level, relies on a proximal Jacobian ADMM scheme and requires the exchange of optimal power injections from the higher level in addition to price signals.
This increases coordination complexity and reduces transparency, as participants are required to incorporate algorithm-specific coupling variables rather than responding solely to economically interpretable price signals. Such a coordination interface contradicts the principle of limited information sharing and undermines the autonomy of market participants, as their local optimization problems must be adapted to account for external algorithmic variables.

Among the methods used for two-level coordination in the literature, some approaches enhance convergence properties through the exchange of multiple variables, such as the previously mentioned ADMM or for example the surrogate Lagrangian relaxation method~[\citenum{Bragin2022}] that utilizes multiple types of Lagrangian multipliers, including nodal shadow prices, interface coupling multipliers, and penalty-based multipliers. 
However, when simplicity and transparency are prioritized, methods based solely on marginal pricing, such as~[\citenum{Andrianesis}], represent a better choice as they maintain decentralized operation with minimal information exchange.
The approach proposed in~[\citenum{Andrianesis}] focuses on the coordination between the DSO and DERs, while treating transmission-level prices as exogenous, thereby decoupling distribution-level power decisions from transmission-level price formation.
Although such a two-level marginal pricing framework can be naturally embedded into integrated transmission–distribution system models, doing so would require a tighter coupling between transmission and distribution system formulations and a broader exchange of information across system levels.
In this work, we extend this approach to a three-level setting by explicitly introducing the TSO as a separate coordination level. 
This enables endogenous power-price coupling, in contrast to sequential coordination schemes, without requiring integrated transmission-distribution optimization, while preserving the benefits of marginal pricing coordination through a novel three-level single-loop iterative framework.
In the proposed context, coordination is understood as a structured interaction among autonomous system levels that independently optimize their objectives while aligning operational decisions through limited information exchange, rather than through centralized optimization, which is consistent with the interpretation adopted in recent literature on hierarchical power system coordination~[\citenum{Muhindo2025}].
Under this interpretation, coordination leads to a consensus understood in a game-theoretic sense as an equilibrium in which, for a given set of exchanged variables, all system levels solve their respective optimization problems optimally and have no incentive to unilaterally deviate.

The main aim of this work is to demonstrate the potential benefits and analyze the performance of the proposed TSO-DSOs-DERAs coordination approach.
Such an approach ultimately enables more efficient integration of distributed resources.
The contributions of this work are listed as follows:
\begin{itemize}[label=--, leftmargin=*, nosep]
    \item
    A novel iterative single-loop price-based coordination framework of the transmission system, distribution systems and aggregated DERs is proposed, enabling three-level decentralized optimization with limited information sharing for multi-interval power scheduling.
    \item 
    The method extends a well-scalable two-level marginal pricing scheme from [\citenum{Andrianesis}] to a three-level TSO–DSOs–DERAs coordination framework by explicitly introducing the TSO as a separate coordination level, while preserving convergence guarantees under limited information exchange.
    \item
    Unlike prior multi-level iterative approaches such as [\citenum{Ullah}], the proposed method limits the exchanged information to top-down price signals only, which are economically interpretable and transparent, thereby aligning with existing market practices and facilitating regulatory acceptance.
    \item 
    Numerical studies demonstrate that the proposed coordination achieves outcomes very close to those obtained with a centralized formulation, while providing clear benefits over sequential pricing-based coordination with fixed transmission system's energy prices in terms of cost reduction and peak shaving.
\end{itemize}

The remainder of the paper is as follows. Section~\ref{s:method} describes the proposed approach.
Section~\ref{s:results} presents the experimental setup, the results obtained, and their discussion.
Section~\ref{s:discussion} discusses the key assumptions and limitations of the proposed approach.
Section~\ref{s:conclusion} concludes the work.

\section{Proposed Method} \label{s:method}

\begin{figure}[ht]
    \centering
    \includegraphics[trim=2.9cm 18.5cm 3cm 2cm, clip, width=3.5in]{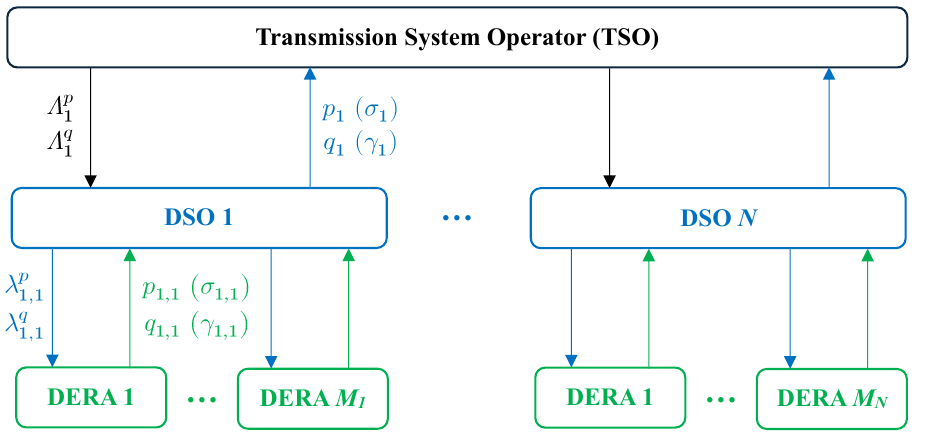} 
    \caption{{Overview scheme of the shared data between TSO, DSOs and DERAs (presented variables are vectors containing the values for all time intervals).}}
    \label{fig:exchange}
\end{figure}

An energy market with multiple time intervals is considered, involving the TSO, DSOs, and DER aggregators. For clarity and focus, the formulation is presented for a day-ahead market with a one-day horizon and hourly time intervals; however, we note that the multi-period coordination in this paper is implemented under the perfect foresight conditions, i.e., we do not default to receding horizon planning. Our approach can be naturally extended to market settings with receding planning and shorter decision intervals, such as intra-day markets.
The goal is to determine the power transfer schedule between the transmission system, distribution systems and DERAs for each time interval with the energy prices corresponding to it.
It is achieved with only limited information sharing between adjacent levels of the hierarchy using the iterative coordination approach.
The higher level entities provide the price values ($\Lambda^{p/q}$, $\lambda^{p/q}$) for all time intervals, while the lower level entities respond with the desired power injection schedule ($p$, $q$).
Additionally, the schedules can be supplemented with uncertainty information to enable a probabilistic formulation.
The example of such a formulation, considered in this work, is the well-scalable AC chance-constrained optimal power flow (AC CC-OPF) proposed in [\citenum{Lubin}].
In that approach, the uncertainty of the active power is described by the standard deviation $\sigma$, while the reactive power deviations are assumed to be proportional to the active power deviations with a ratio $\gamma$.
The overview scheme of the shared information is presented in Fig.~\ref{fig:exchange}.

The following subsections describe the problem formulations of the TSO and DSO, the coordination algorithm, and its convergence guarantee. The proposed coordination framework does not impose any specific requirements on the optimization method applied by DERAs; therefore, this aspect is not detailed in this work.
For completeness, a representative non-convex DER aggregator formulation used in the numerical experiments is provided in Appendix~A.

\subsection{TSO Optimal Power Flow Problem}

TSO managed the power schedules of the large-scale power generators ($p_g$, $q_g$) based on their cost functions ($f^g$) and constraints.
Other entities demand a certain level of power which needs to be satisfied.
Among them, there are both the entities whose power demand is not impacted by the price values, such as loads and the inactive distribution networks ($P^L$, $Q^L$), and the ones whose power demand depends directly on the prices, e.g., active distribution networks managed by DSOs ($P^D$, $Q^D$).
For solving the TSO's optimization problem for a particular iteration, these power injections are considered as parameters.
The objective of TSO is to minimize the cost of power generation.
The power balance and power flow equation are formulated as AC optimal power flow (AC OPF).
The power limits of the generators and grid operating limits, that are power flow and voltage limits, are considered as strict constraints.
As there are no temporally dependent constraints, the problem can be solved separately for each time interval. For simplicity, the time index $t$ is therefore omitted in the formulation in this section.
Nevertheless, it is worth noting that even if such intertemporal constraints were present, the proposed approach would remain applicable.
The nonlinear deterministic formulation is formulated as follows.
\begin{subequations} \label{eq:TSO}
    \begin{align}
        \min_{p, q} \sum_i f^g_i (p^g_i, q^g_i)\\
        \text{s.t.} \quad \Lambda^p_i: p^g_i - P_i^D - P_i^L= v_i^2 G_i^{sh} +\sum_{j: \ i\rightarrow j} p_{ij} \quad \forall i\\
        \Lambda^q_i: q^g_i -Q_i^D -Q_i^L =-v_i^2 B_i^{sh}  +\sum_{j: \ i\rightarrow j} q_{ij} \quad \forall i\\
        p_{ij}=v_i^2 G_{ij} - v_i v_j (G_{ij}\cos{\theta_{ij}} + B_{ij} \sin{\theta_{ij}}) \quad \forall i,j\\
        q_{ij}=-v_i^2 B_{ij} - v_i v_j (G_{ij}\sin{\theta_{ij}} - B_{ij} \cos{\theta_{ij}}) \quad \forall i,j\\
        \underline{P^g_i} \leq p^g_i \leq \overline{P^g_i}, \quad 
        \underline{Q^g_i} \leq q^g_i \leq \overline{Q^g_i} \quad \forall i\\
        p_{ij}^2 + q_{ij}^2 \leq \overline{S}_{ij}^2 \quad \forall i,j\\
        \underline{V_i} \leq v_i \leq \overline{V_i} \quad \forall i
    \end{align}
\end{subequations}

The problem (\ref{eq:TSO}) is linearized and reformulated to the chance-constrained form (AC CC-OPF) for which the power injection uncertainties are taken into account as proposed in~[\citenum{Lubin}].

\subsection{DSO Power Flow Problem} \label{s:DSO}

Each DSO determines the power transfer from the transmission system and DLMPs for each time interval.
As the problem is time separable, the time index is omitted in the presented formulation.

All power injections in the distribution system, e.g., from DERAs ($P^A$, $Q^A$), are assumed to be controlled indirectly by determining distribution locational marginal prices (DLMPs), $\lambda^p$ and $\lambda^q$, and no additional decision variables related to power scheduling are introduced at the distribution level.
It is assumed that the DSO can achieve the nominal slack bus voltage for any transmission-side voltage within the admissible operating limits, e.g., by adjusting the taps of the substation transformer.
As a consequence, the power flow calculations have only one feasible solution which does not depend on the TSO problem.
The deterministic power flow problem is formulated as a Distribution Flow (DistFlow) problem as follows.
\begin{subequations}
    \begin{align}
        p_i+P_i^A = R_i l_i + P_i^L + \sum_{j: \ i\rightarrow j} p_{j},  \quad  
        q_i+Q_i^A = X_i l_i + Q_i^L + \sum_{j: \ i\rightarrow j} q_{j} \quad \forall i\\
        v_j^2 l_i = p_i^2+q_i^2, \quad
        v_i^2 = v_j^2 - 2 \cdot R_i p_i - 2 \cdot X_i q_i + Z_i^2 l_i \quad
        \forall i, \ j: j \rightarrow i
    \end{align}

Based on the its solution, the violation variables are calculated as follows.
    \begin{align}
        \psi_i = \max\{0; \ \sqrt{p_i^2+q_i^2} - \overline{S_i}\}, \quad \delta_i = \max\{0; \ \underline{V_i} - v_i; \ v_i-\overline{V_i} \} \quad  \forall i
    \end{align}

Similarly as in the case of the transmission system, to take into account the uncertainty of the power injections, the presented set of equations is reformulated to the chance-constrained form, based on the approach presented in [\citenum{Lubin}].

To simplify the notation, this set of power flow equations is denoted as $H$, its variables as $y$ and the solution as $y^*=\arg\{H(y)=0\}$. 

The objective function aims to minimize the cost of the power transfer from the transmission system ($p_0$, $q_0$) and the penalty function ($f_{pen}$) for the grid operating limits violations.
\begin{align}
    F_D(y) = \Lambda^p \cdot p_0 + \Lambda^q \cdot q_0 + f_{pen}(\psi, \delta)
\end{align}

Then DLMPs for bus $i$ are determined as a sensitivity parameters of the objective function as follows.
\begin{align}
    \lambda_i^p = \frac{\partial F_D(y^*)}{\partial p_i}, \quad 
    \lambda_i^q = \frac{\partial F_D(y^*)}{\partial q_i}
\end{align}
\end{subequations}

\subsection{Coordination Algorithm}

It is assumed that there is a market operator that mediates the transfer of data between entities.
The coordination between the adjacent levels is performed based on the (distribution) locational marginal price values.

\begin{figure}[ht]
    \centering
    \includegraphics[trim=3cm 8cm 3cm 2.5cm, clip, width=3.5in]{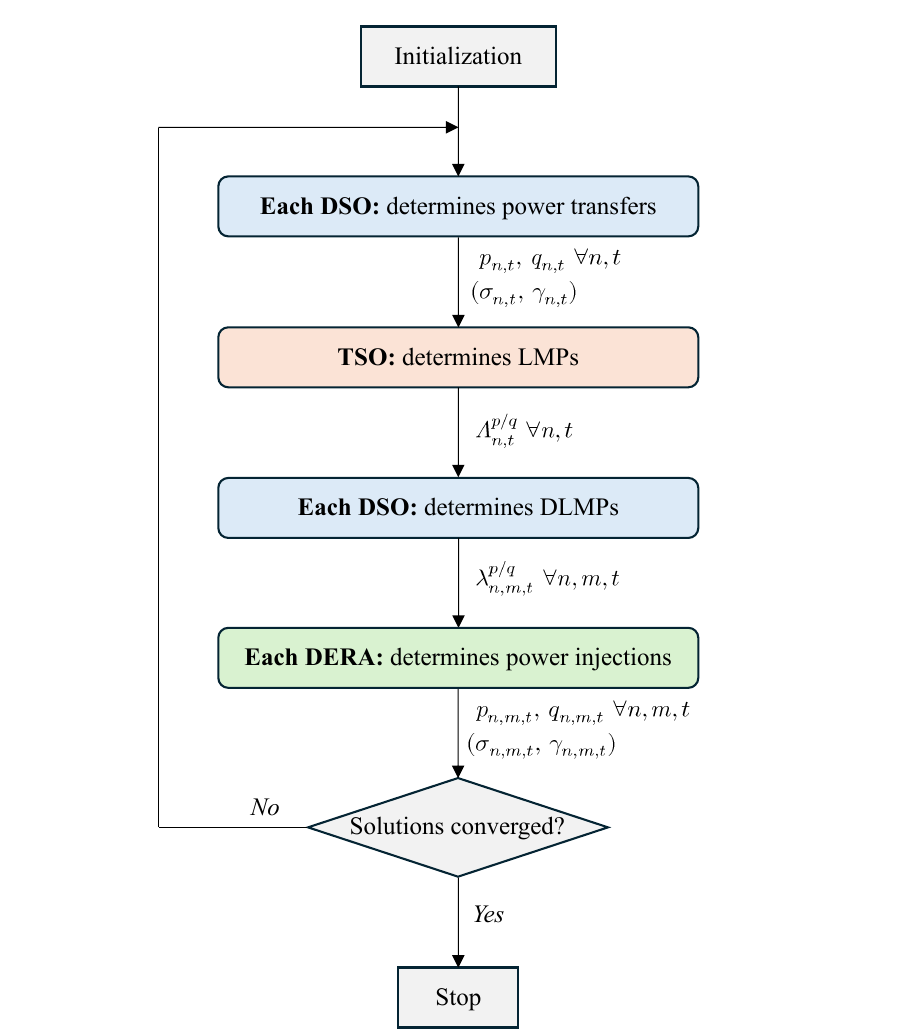}
    \caption{{The flowchart of the TSO-DSO-DERAs coordination algorithm.}}
    \label{fig:algorithm}
\end{figure}

The algorithm flowchart is presented in Fig.~\ref{fig:algorithm}.
The calculations are performed for all time intervals within time horizon.
Distributed subproblems of DSOs and DERAs are solved in parallel, and their results are synchronized before passing to the next coordination step.
The process is initialized based on the predictions of the load demand and renewable energy generation, without taking into account the flexibility of the resources.
For each iteration, each DSO determines the demanded power transfers from the transmission system ($p_{n,t}$, $q_{n,t}$) based on the DERAs solutions from the previous iteration.
Based on these power values, TSO solves its optimization problem and determines LMPs ($\Lambda_{n,t}^{p/q}$) for each bus.
Using these LMPs each DSO updates DLMPs ($\lambda_{n,m,t}^{p/q}$).
Based on the DLMPs, each DERA solves its optimization problems and provides the power injection values ($p_{n,m,t}$, $q_{n,m,t}$).
Considering the probabilistic approach, the power schedules of the DERAs and DSOs are supplemented by the uncertainty variables ($\sigma$, $\gamma$).
The termination condition is defined based on the price differences between consecutive iterations.

\subsection{Convergence Properties}

The proposed single-loop three-level coordination does not introduce new convergence mechanisms. Instead, it builds on two-level proximal coordination framework in [\citenum{Andrianesis}], from which it draws convergence guarantees, and expands this mechanism to a three-level setting under non-restrictive assumptions.
These assumptions concern the intermediate (distribution) level, which is modeled as a non-decision-making network layer whose power flow is decoupled from the transmission system optimization.
Specifically, this is ensured by indirect control of all power injections through DLMPs, the absence of additional decision variables at the distribution level, and the maintenance of the nominal slack-bus voltage for any transmission-side voltage within the admissible operating limits (as described in Section~\ref{s:DSO}).

The integrated optimization problem of all levels of the system can be represented by the following general formulation.
\begin{subequations}
    \begin{align}
        \min_{x,y,z} f_T(x) + f_D(y) + f_A(z) \\
        \text{s.t.} \quad \quad \quad \ Ax+By=a \label{eq:co_1}\\
        Cy+Dz=b \label{eq:co_2}\\
        x \in \mathbf{X}, \quad y \in \mathbf{Y}, \quad z \in \mathbf{Z} \label{eq:co_3}
    \end{align}
\end{subequations}

In the presented formulation $x$, $y$ and $z$ are variable vectors of the transmission system, distribution systems, and DERAs, respectively.
$f_T(x)$, $f_D(y)$ and $f_A(z)$ are their respective objective functions, which do not include the cost of the power transfer between the levels.
Equations (\ref{eq:co_1}) and (\ref{eq:co_2}) represent power balance constraints for the connection points between the levels, as the vectors $By$ and $Dz$ contain the active and reactive power transfer values between the transmission and distribution systems and the distribution systems and DERAs, respectively.
Other constraints of the transmission, distribution systems and DERAs are represented in equation (\ref{eq:co_3}) by the sets $\mathbf{X}$, $\mathbf{Y}$ and $\mathbf{Z}$, respectively.

This integrated problem can be separable between two levels: the power system level (integrated transmission and distribution system problems) and DERAs level and solved iteratively using the approach proposed in [\citenum{Andrianesis}].
The formulation for iteration $k$ can be presented as follows.
\begin{subequations}
    \begin{align}
        \min_{x,y \in \mathbf{XY}} f_T(x) + f_D(y) \ \text{s.t.} \  \lambda^k: Cy+Dz^{k-1}=b\\
        z^{k} = \text{arg} \min_{z \in \mathbf{Z}} \{f_A(z)+\lambda^k Dz + \Psi^k\}
    \end{align}
\end{subequations}

The set $\mathbf{XY}$ represents all constraints of the power system apart from the ones related to the DERAs power injections.
The co-optimization term $\Psi$ which penalizes the difference of the power injection between consecutive iterations, with the penalization parameter $\rho$, is defined as follows.
\begin{equation}
    \Psi^k = \rho^k \cdot ||Dz-Dz^{k-1}||^2
\end{equation}

This two-level approach enjoys convergence properties of proximal algorithms, including cost descent behavior for sufficiently large positive scalar parameter $\rho$, as discussed in [\citenum{Andrianesis}].

As for the two-level case with the integrated optimization problem of the transmission and distribution systems, the convergence properties of the coordination approach between the power system and DERAs are established in the literature; the following considerations aim to demonstrate that the separation of the transmission and distribution problems in the proposed coordination approach leads to the same results in every iteration as in the two-level case.

The equations describing the proposed three-level coordination with four main calculation steps are formulated as follows.
\begin{subequations}
    \begin{align}
        y^k = \text{arg}\{y \in \mathbf{Y}: \ Cy+Dz^{k-1}=b\} \label{eq:alg_a}\\
        x^k=\text{arg} \min_{x \in \mathbf{X}} f_T(x) \quad
        \text{s.t.} \quad \Lambda^k: \ Ax+By^{k}=a \label{eq:alg_b}\\
        \lambda^k = \frac{\partial (f_D(y^k)+\Lambda^k By^k)}{\partial p}, \quad 
        p=b-Cy^k \label{eq:alg_c}\\
        z^{k} = \text{arg} \min_{z \in \mathbf{Z}} \{f_A(z)+\lambda^k Dz + \Psi^k\} \label{eq:alg_d}
    \end{align}
\end{subequations}

The proposed three-level coordination and the two-level approach yield identical solutions in each iteration, provided that the power flow results and the calculated DLMP values are the same.
As discussed in Section~\ref{s:DSO}, the DSO power flow calculations (represented by~\eqref{eq:alg_a}) are independent of the TSO results and can therefore be performed separately in advance.
Given that the distribution system power flows are identical, the TSO solution \eqref{eq:alg_b} also matches that of the two-level approach.
The analysis concerning the DLMP values in~\eqref{eq:alg_c} follows below.

Active and reactive power DLMPs for the distribution system $n$ and bus $m$ are equal to the sensitivity of the objective function in respect to the power injection $p_{n,m}$ and $q_{n,m}$, respectively.
Taking as an example the active power DLMP for the integrated problem, it can be described as follows.
\begin{subequations}
    \begin{align}
        \lambda^{int,p}_{n,m}= \frac{\partial (f_T+f_D)}{\partial p_{n,m}}
        = \frac{\partial f_{T}}{\partial p_{n,m}} + \frac{\partial f_{D}}{\partial p_{n,m}} \label{eq:lambda_int}
    \end{align}

For the separated problem, the objective of the distribution system includes the cost of the power transfer ($p_n$, $q_n$) from the transmission system bases on LMPs, so DLMPs are equal as follows.
    \begin{align}
        \lambda^p_{n,m}= \frac{\partial (\Lambda^p_n \cdot p_n + \Lambda^q_n \cdot q_n + f_{D})}{\partial p_{n,m}} 
        =  \Lambda^p_n \cdot \frac{\partial p_n }{\partial p_{n,m}} 
        + \Lambda^q_n \cdot \frac{\partial q_n }{\partial p_{n,m}}
        + \frac{\partial f_{D}}{\partial p_{n,m}}
    \end{align}
As LMPs are determined as sensitivities of the transmission system objective function in respect to the distribution system power transfer, this equation can be presented as follows.
    \begin{align}
        \lambda^p_{n,m}= \frac{\partial f_T}{\partial p_n} \cdot \frac{\partial p_n }{\partial p_{n,m}} 
        + \frac{\partial f_T}{\partial q_n} \cdot \frac{\partial q_n }{\partial p_{n,m}}
        + \frac{\partial f_{D}}{\partial p_{n,m}} \label{eq:lambda_sep}
    \end{align}
Based on the derivative chain rule it can be stated:
\begin{align}
    \frac{\partial f_T}{\partial p_{n,m}} =  
    \frac{\partial f_T}{\partial p_n} \cdot \frac{\partial p_n }{\partial p_{n,m}} 
    + \frac{\partial f_T}{\partial q_n} \cdot \frac{\partial q_n }{\partial p_{n,m}}
\end{align}

Then, the DLMPs for the integrated (\ref{eq:lambda_int}) and separated (\ref{eq:lambda_sep}) power system problems are equal to each other.
\end{subequations}

As the solutions for the integrated and separated power system problems are the same for each iteration, the convergence guarantees for the two-level problem are applicable for the proposed three-level approach.

Accordingly, the proposed single-loop extension does not increase the number of coordination iterations compared to the two-level formulation based on an integrated transmission-distribution system model.
The main additional overhead arises from separating the network calculations into three computational steps, namely the determination of power transfers by DSOs, LMPs by the TSO, and DLMPs by DSOs, which requires only limited exchange of price signals and aggregated power schedules across coordination levels.
Importantly, the dominant computational burden remains associated with the TSO OPF, the DSO network calculations, and the DERAs’ local optimization problems, which are already present in the integrated two-level approach. Consequently, the additional overhead introduced by the proposed separation is marginal relative to the overall computational effort.

\section{Experimental Setup and Results} \label{s:results}

\subsection{Experimental Setup}
The IEEE 9-bus transmission system is considered. The line, bus, and generation data are based on [\citenum{IEEE9}].
The generators are assumed to operate continuously and therefore the no-load cost is neglected.
As the load of this system (at buses 5, 7, and 9 with nominal active power equals 90 MW, 100 MW, and 125 MW respectively), the modified IEEE 33-bus distribution systems [\citenum{IEEE33}] are considered, which are rescaled in such a way that their nominal active power demand matches the nominal load of the transmission system.
The line power limits for the distribution system, not specified in used dataset, are assumed as 30\% greater than the apparent power flow for the nominal load value.
At each bus of each distribution system, DERA is considered.
Each DERA manages RES, specifically PV units, with power rating equals 30\% of the nominal load value for this bus, FL (shiftable by one hour) being 20\% of the load, and BESS with power rating in range 0-30\% of the nominal load value with three-hour energy storage capability at full power.
The power profiles of the load demand and PV generation are based on~[\citenum{Power}].
The minimum value of power factor ($\underline{PF}$) for PV unit is equal 0.8, the efficiency ($\eta$) of the converter of PV and battery energy storage system (BESS) is equal 90\%, and the maximum value of the power consumption of flexible loads for specific time interval ($\overline{P_{f,t}}$) is two times greater than the default power value ($P_{f,t}$).
The parameters related to the AC CC-OPF include a standard deviation of PV generation equal to 20\% of the forecast value, which is assumed to be normally distributed, and a probability level of 1\% for all chance constraints, namely generator and line power limits as well as bus voltage limits.
For the presented cases, the optimal solutions follow the grid operating limits of the distribution systems, so the values of the penalty coefficients ($\kappa_S$, $\kappa_V$) have no impact on the results.

\subsection{Results Comparison with the Reference Approaches}

\subsubsection{Reference Approaches}
As the first reference approach, a centralized formulation is considered, which provides the optimal solution obtained under full information sharing across all system levels. This reference is used to evaluate potential deviations introduced by the proposed distributed coordination scheme.

The proposed coordination method does not aim to improve the quality of the solution itself, but rather to achieve comparable results under stricter architectural constraints, specifically, with limited information sharing between system levels and a simpler, more transparent coordination structure. 
For this reason, comparisons with approaches involving broader information exchange are limited to the fully centralized formulation, which represents the extreme case of complete system integration.

As the main reference approach, a commonly used sequential coordination approach is used as a reference, in which the TSO determines LMPs based on predicted net DSO demand, and the DSO–DERA coordination proceeds with these fixed LMPs.
This approach represents a fair comparison, as it relies on consistent architectural assumptions, namely a coordination framework with a single iterative loop, in which only price signals are passed top-down and power values are sent bottom-up across all three~levels.

\subsubsection{Results Comparison}
\begin{figure}[ht]
    \centering
    \includegraphics[trim=3.5cm 9cm 4.5cm 9cm, clip, width=0.4\linewidth]{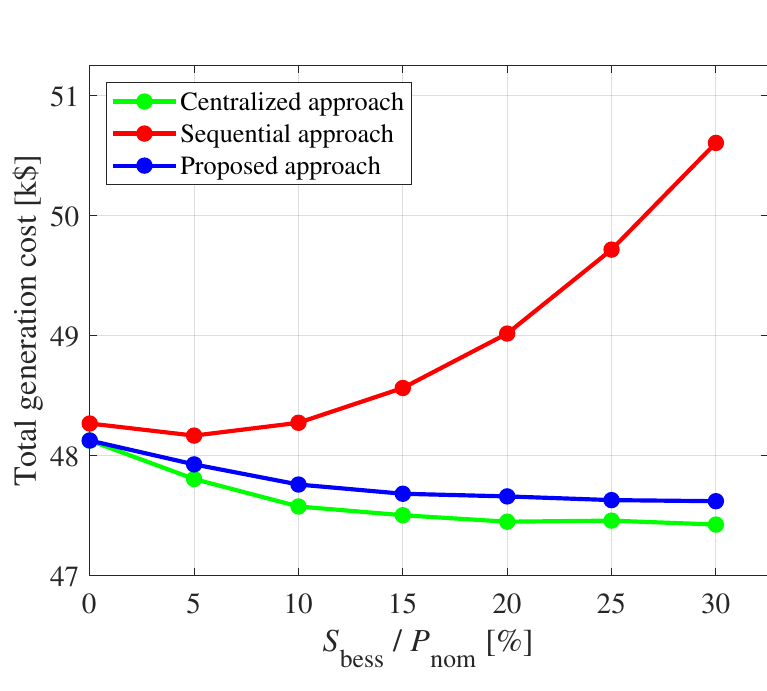}
    \caption{{Total generation cost for different BESS power ratings.}} 
    \label{fig:cost}
\end{figure}

\begin{figure}[ht]
    \centering
    \includegraphics[trim=3.5cm 8cm 4.5cm 9cm, clip, width=0.4\linewidth]{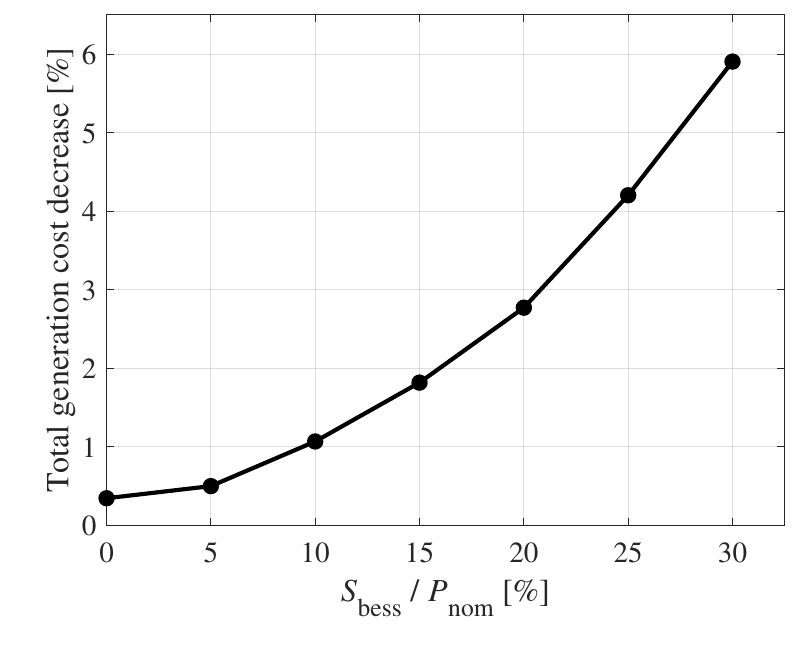}
    \caption{{Generation cost's relative decrease in relation to the sequential reference approach.}} 
    \label{fig:cost_diff}
\end{figure}

\begin{figure}[ht]
    \centering
    \includegraphics[trim=3.5cm 8cm 4.5cm 9cm, clip, width=0.4\linewidth]{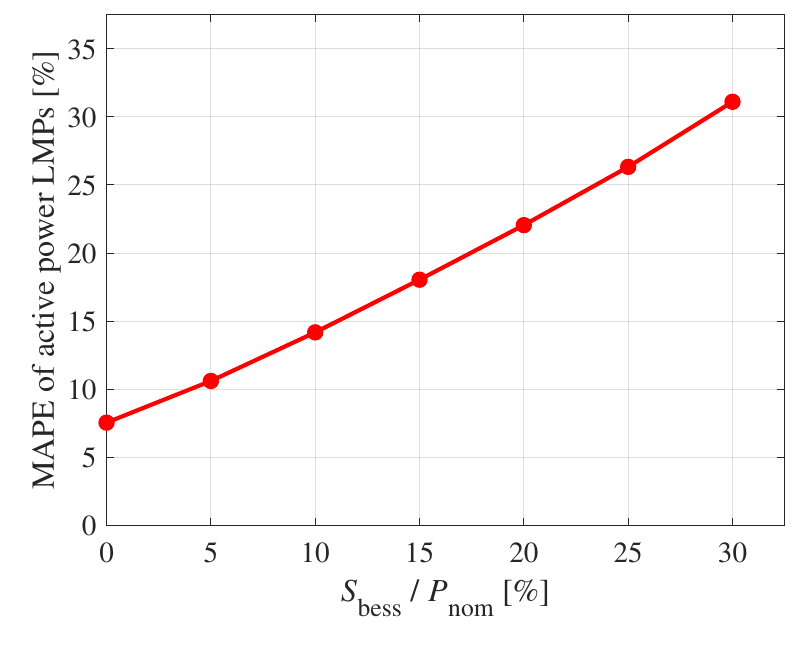}
    \caption{{MAPE ($\varepsilon$) of the sequential reference approach's active power LMPs.}}
    \label{fig:price_error}
\end{figure}

\begin{figure}[ht]
    \centering
    \includegraphics[trim=3.5cm 9cm 4.5cm 9cm, clip, width=0.4\linewidth]{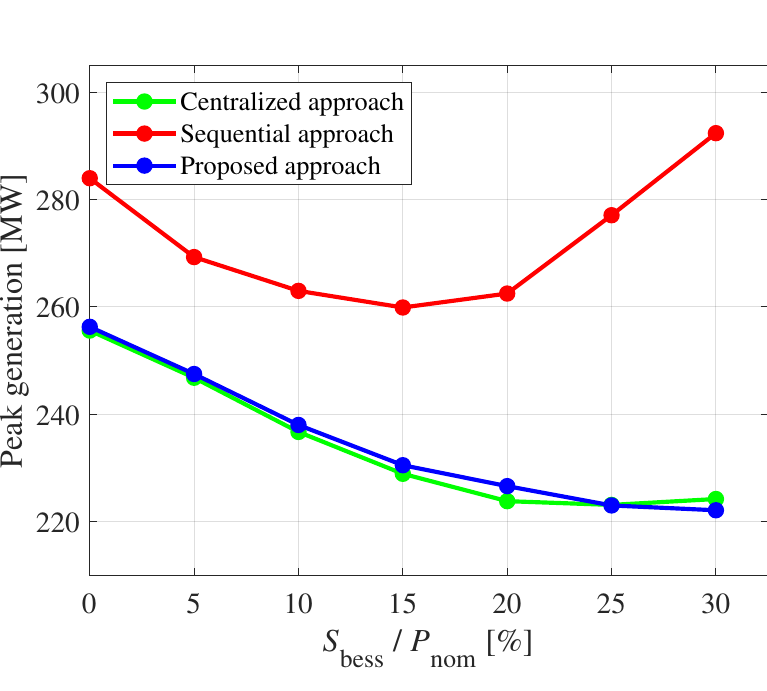}
    \caption{{Peak generation for different relative BESS power ratings.}}
    \label{fig:peak}
\end{figure}

\begin{figure}[ht]
    \centering
    \includegraphics[trim=3.5cm 8cm 4.5cm 9cm, clip, width=0.4\linewidth]{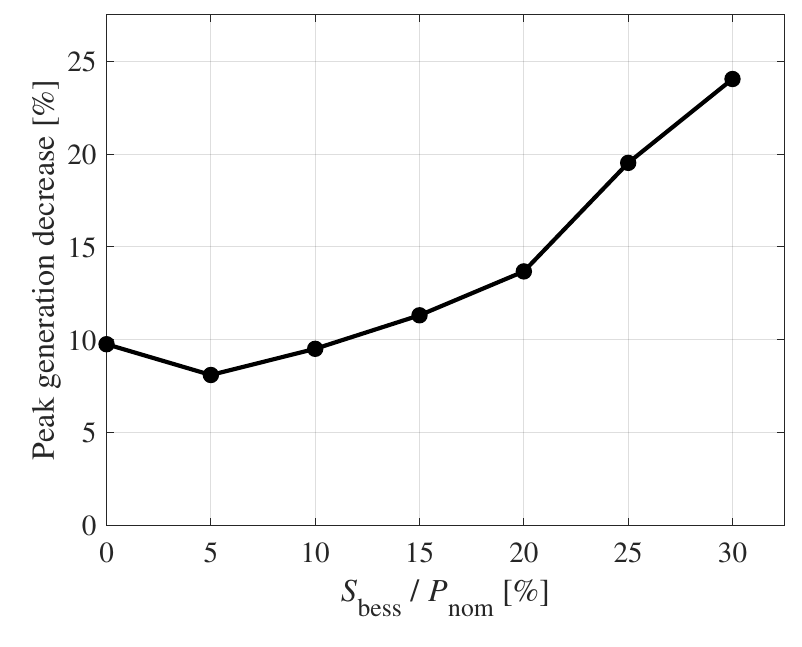}
    \caption{{Peak generation's relative decrease in relation to the sequential reference approach.}}
    \label{fig:peak_diff}
\end{figure}

Primarily, the results are compared in terms of the economic performance.
As the objectives of DERAs and DSOs do not have any cost apart from the payments to other levels, the total cost for the whole system is equal to the generation cost.
The results in terms of the one-day expected generation cost ($C$) for different levels of relative (to the nominal load values) BESS power ratings ($S_{bess}$), i.e. different levels of the DERAs flexibility, are compared in Fig.~\ref{fig:cost} and \ref{fig:cost_diff}.
The mean average percentage error (MAPE, $\varepsilon$) of the active power LMPs for the sequential reference approach ($\Lambda^p_{ref}$) is presented in Fig.~\ref{fig:price_error}.
It is calculated as a relative difference between the used fixed values of LMPs and the actual values calculated for the final solution.
Moreover, as the metric which indirectly corresponds to the system operating limits, the peak generation of the of the transmission system's conventional generators ($P^{max}_g$) is used.
These values are compared in Fig.~\ref{fig:peak} and \ref{fig:peak_diff}.

Across all considered scenarios, the generation costs obtained with the proposed coordination approach remain very close to the centralized reference solution, with deviations not exceeding 0.5\%, indicating that the distributed formulation introduces only minor optimality losses.
Similarly, the peak generation levels are also very close in both cases.

For a relatively small level of DERAs flexibility, i.e., without BESS (with only FL), the difference between the generation cost for the proposed and sequential reference method is very small (0.3\%).
The relative error of LMPs for the sequential method in that case is over 7\%.
Increasing the level of DERAs flexibility, by increasing the relative BESS power rating, leads to a faster than linear increase of the relative decrease of the generation cost of the proposed method in relation to the sequential one.
For the highest considered value of relative BESS power rating equals 30\%, this difference is equal nearly 6\%.
It is worth noticing that the reference generation cost is slightly decreasing for the relatively low level of flexibility, and increasing significantly for relatively high levels.
In the case of the proposed method, the generation cost with increasing the flexibility level is decreasing at a diminishing rate.
The relative error of LMPs for the sequential method is roughly linearly increasing, exceeding 31\% for the highest considered BESS power rating.

The peak generation in the case without BESS is almost 10\% lower for the proposed method in comparison to the sequential reference method.
The proposed method leads to a decrease of the peak generation in the whole considered range of BESS power ratings.
The peak generation of the sequential method is decreasing for the BESS relative power ratings lower than 20\%, but for the higher values, it is increasing significantly, exceeding the initial value.
As a consequence, the relative difference of the peak generation for the proposed and sequential method is increasing for the BESS relative power rating greater than 10\%, reaching 24\%.

\begin{figure}[ht]
    \centering
    \includegraphics[trim=3.5cm 9cm 4.5cm 9cm, clip, width=0.4\linewidth]{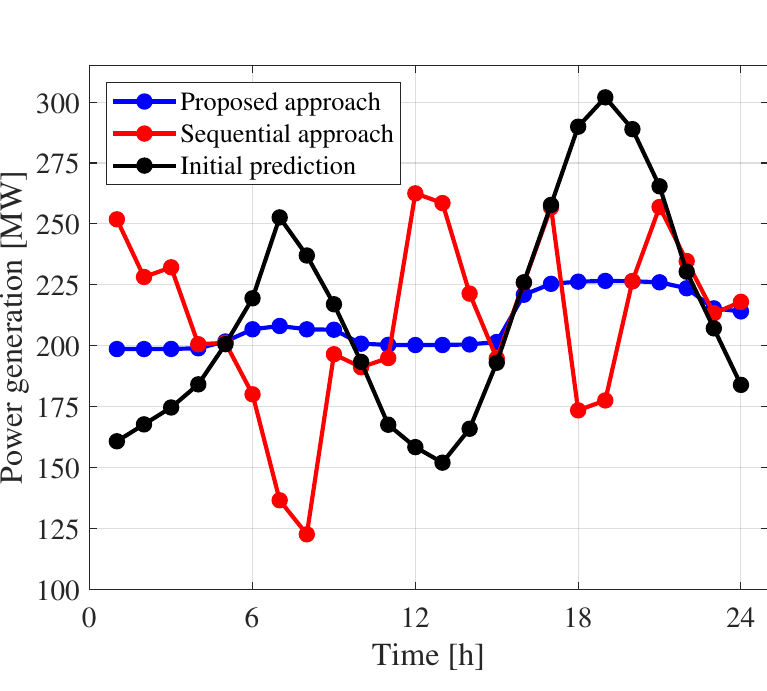}
    \caption{{Exemplary power generation profiles (for $S_{bess}=20\%$).}}
    \label{fig:power}
\end{figure}

This significant difference between the results of the proposed and sequential reference approach for the relatively high level of DERAs flexibility can be explained as follows.
The three-level coordination with simultaneous LMP update tends to the solution which minimizes the total generation cost.
Assuming constant total net demand and quadratic cost functions of power generation, the optimal solution tends to approach the average value of the power demand for all time intervals.
As the algorithms tend to minimize the power transfer for the time with lowest prices, and maximize it for the time with highest prices, the approach with fixed LMPs tends to increase the power demand for the trough of the initial power prediction and vice versa.
This tendency leads to a decrease of the generation cost and peak generation as long as the flexibility is not enough to inverse the time of the peaks and troughs.
However, for the relatively high levels of flexibility, increasing the flexibility level leads to an increase of deviations from average power values (in the opposite direction in comparison to the power prediction) and, as a consequence, increases the generation cost as well as peak generation.
The exemplary power generation for the case with relatively high DERAs flexibility is presented in Fig.~\ref{fig:power}.

\subsection{Computational Performance and Scalability}

To evaluate computational performance and scalability in addition to the previously described main test system, a larger system is also considered. For this purpose, the IEEE 57-bus transmission system is used [\citenum{IEEE57}], with extended voltage limits in the range of 0.90–1.10 pu. Twelve load buses with the highest active power demand are replaced with a rescaled IEEE 33-bus distribution system each, similarly to the configuration adopted in the main test system.

In this work, computational performance is primarily assessed in terms of the number of coordination iterations required for convergence, since the wall-clock time of individual iterations depends on the model fidelity and solver implementation at each system level. Therefore, the convergence behavior is illustrated through the DLMP changes over consecutive iterations.

\begin{figure}[ht]
    \centering
    \includegraphics[trim=3cm 9cm 4cm 9cm, clip, width=0.4\linewidth]{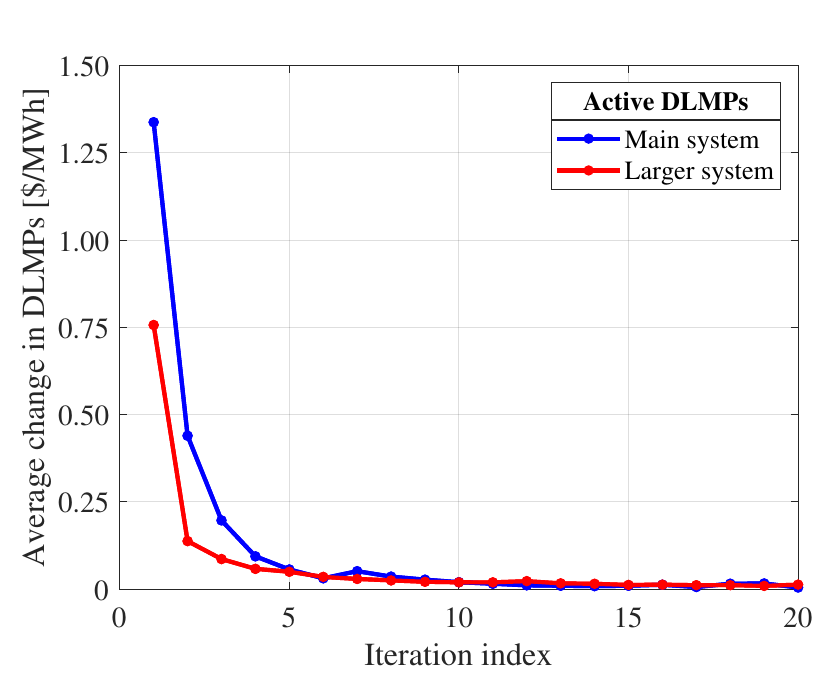}
    \includegraphics[trim=3cm 9cm 4cm 9cm, clip, width=0.4\linewidth]{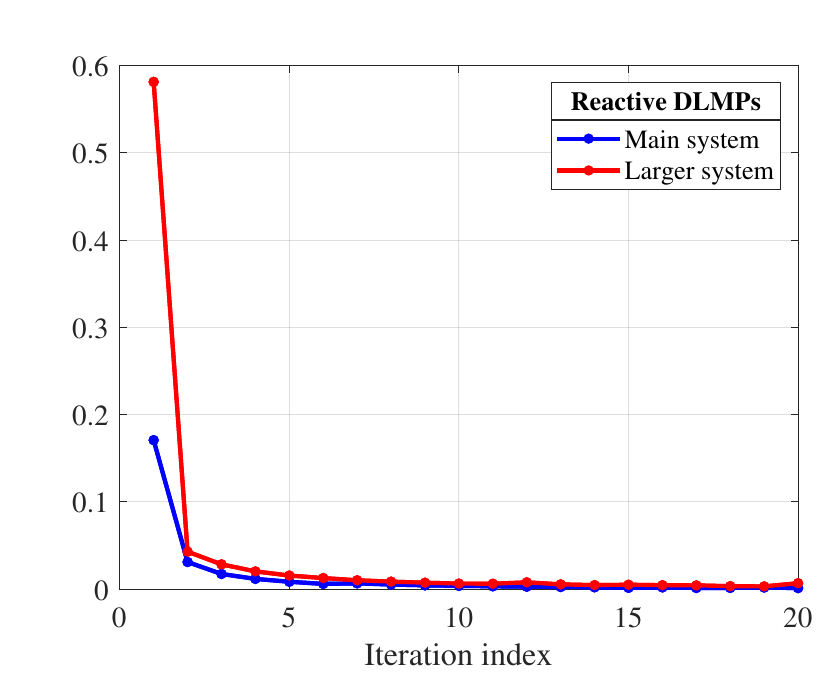}
    \caption{{Average absolute changes in active and reactive DLMPs between consecutive iterations for the main and larger systems ($S_{bess}=20\%$).}}
    \label{fig:convergence}
\end{figure}

Fig.~\ref{fig:convergence} presents the average absolute changes in active and reactive DLMPs for the case with $S_{bess}=20\%$. The penalization parameter $\rho$ is selected such that a change in the DERA power schedule equal to 1\% of the nominal load of the distribution system results in a penalty of 0.3\$. The average converged DLMP values for the main system are 18.09~\$/MWh and 0.59~\$/MVAh for active and reactive power, respectively. For the larger system, the corresponding values are 44.09~\$/MWh and 1.45~\$/MVAh.

The results indicate that DLMP variations become relatively small after only a few iterations and decrease to negligible levels after a limited number of additional iterations for both systems, suggesting broadly similar convergence behavior.

While the number of coordination iterations is the primary performance indicator in this work, the corresponding wall-clock computation times are reported for completeness and illustrative purposes.
The proposed coordination framework is designed for a distributed implementation in which DSO and DERA subproblems are solved independently and can be executed in parallel. As a result, the wall-clock time of a single coordination iteration is determined by the slowest individual subproblem rather than by the total number of participating DSOs or DERAs.

All simulations were performed in MATLAB~2024b on a system equipped with a 16-core 4.3~GHz processor and 48~GB of RAM.
In each coordination iteration, individual DSO and DERA subproblems were solved on the order of a few seconds.
For the main test system, the TSO problem was also solved within a few seconds, whereas for the larger transmission system the solution time increased to approximately one minute.

\section{Discussion} \label{s:discussion}

The assumptions regarding the intermediate (distribution) level are introduced in Section 2.2, where the distribution system is modeled as a non-decision-making network layer, excluding additional decision variables such as network reconfiguration or slack bus voltage regulation. Under these assumptions, the proposed three-level coordination framework is equivalent, on an iteration-by-iteration basis, to the two-level approach with an integrated transmission-distribution system formulation. Even if additional decision variables were introduced at the distribution level, the same equivalence could still be obtained under further structural assumptions, in particular if the distribution-level objective function is monotonic with respect to the active power transfer from the transmission system. In such cases, the optimal solution of the distribution problem does not depend explicitly on the magnitude of the active LMP signal, for example when the sign of the active LMP is known a priori and reactive LMPs are neglected.

The coupling between reactive LMPs and reactive DLMPs is introduced as a modeling choice to allow for a more general coordination framework rather than as an operational requirement. The proposed algorithm remains applicable when transmission and distribution systems are decoupled in terms of reactive power, consistent with conventional voltage control practices.
From a coordination perspective, this coupling also helps avoid potential boundary mismatches at the transmission–distribution interface that could otherwise lead to suboptimal dispatch decisions.

It should be noted that marginal pricing schemes, including LMPs and DLMPs, are subject to well-known limitations, particularly in the presence of non-convexities and temporal dependencies arising from power system operation, as well as uncertainty modeling approaches~[\citenum{Tan2022}].
In such settings, marginal prices may fail to fully recover costs or provide equilibrium-supporting incentives without additional settlement mechanisms. The proposed coordination framework adopts marginal prices as coordination signals rather than as a complete market-clearing mechanism and does not aim to address these broader pricing challenges, which have been extensively discussed in the literature.

While uncertainty is explicitly considered through a chance-constrained formulation to ensure feasibility under power imbalances and reserve activation, a dedicated analysis of its impact on the results is beyond the scope of this work.

\section{Conclusion} \label{s:conclusion}
The paper proposed a coordination approach for the transmission system, distribution systems, and DERAs. The method is based on a transparent and scalable single-loop iterative mechanism, in which only prices and power schedules are exchanged between adjacent levels.
The calculation for the distribution systems and DERAs is performed in a decentralized way, ensuring good scalability properties of the proposed method.
The algorithm exhibits convergence properties resulting from its proximal-gradient structure. 
The proposed method was compared against a centralized reference formulation, as well as against sequential reference approach with fixed LMPs for DSO-DERA coordination, showing outcomes that remain very close to the centralized solution while significantly improving performance relative to the sequential reference approach.
Simulation results show that higher DER flexibility leads to greater cost and peak reduction benefits under the proposed coordination. 
For example, for 20\% flexible load (shiftable by one hour) and a BESS power rating equal to 20\% of the nominal load value with three-hour energy storage capacity at full power, the total generation cost decreased by about 2.8\%, and the peak generation by~13.7\%. 

\appendix

\section{DERA Optimization Problem} \label{s:appendix}
This appendix provides a representative DER aggregator optimization model used in the numerical experiments.
It is assumed that DERAs aggregate different types of DERs, such as renewable energy sources (RES), battery energy storage systems (BESS) and flexible loads (FL).
Their general goal is to maximize the total reward for total power injections of active and reactive power.
\begin{subequations}
    \begin{align}
        \min_{p,q} \sum_t -(\lambda_t^p \cdot p_t + \lambda_t^q \cdot q_t)\\
        \text{s.t.} \ \forall t: \quad p_t=p_{r,t} + p_{b,t} - p_{f,t},
        \quad q_t=q_{r,t} + q_{b,t} - q_{f,t} \\
        \text{RES, BESS and FL constraints}
    \end{align}
For the RES, the expected available DC power generation, $P_{r,t}$, is known and the power curtailment is possible.
The reactive power values are constrained by the requirement of the minimal power factor $\underline{PF_r}$.
The power losses $\Delta p_{r,t}$, are assumed to be proportional to the apparent power value based on the efficiency $\eta$, which results in a non-convex quadratic equality constraint.
\begin{align}
     0 \leq p_{r,t}^{dc} \leq P_{r,t}, \quad 
     -p_{r,t} \sqrt{1-\underline{PF_r}^2} \leq q_{r,t} \leq p_{r,t} \sqrt{1-\underline{PF_r}^2}\\
     p_{r,t}=p_{r,t}^{dc}-\Delta p_{r,t}, \quad \Delta p_{r,t}^2 = (1 - \eta)^2 \cdot (p_{r,t}^2 +q_{r,t}^2), \quad \Delta p_{r,t} \geq 0
\end{align}
The battery's state of charge $e_t$ is bounded by the operating limits $\underline{E}$ and $\overline{E}$.
The reactive power is constrained by the apparent power limit $\overline{S_b}$.
The power losses $\Delta p_{b,t}$ are calculated in the same way as in case of RES.
\begin{align}
    e_{t+1} = e_t - p_{b,t}^{dc} \cdot \Delta t,  \quad  
    \underline{E} \leq e_t \leq \overline{E}, \quad
    p_{b,t}^2 + q_{b,t}^2 \leq \overline{S_b}^2\\
    p_{b,t}=p_{b,t}^{dc}-\Delta p_{b,t}, \quad \Delta p_{b,t}^2 = (1 - \eta)^2 \cdot (p_{b,t}^2 +q_{b,t}^2), \quad \Delta p_{b,t} \geq 0
\end{align}
It is assumed that the flexible load $P_{f,t}$ can be shifted for one hour, i.e., deferred to the next time interval ($p^d_{t}$) as well as consumed in advance ($p^a_{t}$).
There is a limit of the power consumption for each time interval that equals $\overline{P_{f,t}}$.
Reactive power is calculated by assuming a constant power factor for a specific time interval.
\begin{align}
    p_{f,t} = P_{f,t} - p^d_{t} - p^a_{t} + p^d_{t-1}+p^a_{t+1}, \quad 0 \leq p_{t} \leq \overline{P_{f,t}}\\
    p_t^d \geq 0, \quad  p_t^a \geq 0, \quad p^d_t + p^a_t \leq P_{0,t}, \quad p^a_{1}=0, \quad p^d_{t_{end}}=0\\
    q_{f,t}=(P_{f,t} - p^d_t - p^a_t) \cdot \tan{\varphi_t} + q^d_{t-1} + q^a_{t+1} \quad q^{a/d}_{t} = p^{a/d}_{t} \cdot \tan{\varphi_{t}}
\end{align}
\end{subequations}

\section*{Acknowledgement}
This publication is based upon work supported by King Abdullah University of Science and Technology (KAUST) under Award ORFS-2022-CRG11-5021 and the KAUST Center of Excellence for Renewable Energy and Storage Technologies (CREST), under award \#5937.

\bibliographystyle{elsarticle-num} 
\bibliography{references}

\end{document}